\documentclass[10pt,letterpaper]{article}
\usepackage{opex3}
\usepackage{subfigure}
\usepackage{amsmath}

\begin{document}

\title{Oxides and nitrides as alternative plasmonic materials in the optical range}

\author{Gururaj V. Naik$^1$, Jongbum Kim$^1$ and Alexandra Boltasseva$^{1,2,*}$}

\address{$^1$Birck Nanotechnology center and School of Electrical \& Computer Engineering Purdue University West Lafayette Indiana 47906 USA\\$^2$DTU Fotonik, Technical University of Denmark, Lyngby 2800, Denmark}

\email{*aeb@purdue.edu} 



\begin{abstract}
As alternatives to conventional metals, new plasmonic materials offer many advantages in the rapidly growing fields of plasmonics and metamaterials. These advantages include low intrinsic loss, semiconductor-based design, compatibility with standard nanofabrication processes, tunability, and others. Transparent conducting oxides such as Al:ZnO, Ga:ZnO and indium-tin-oxide (ITO) enable many high-performance metamaterial devices operating in the near-IR. Transition-metal nitrides such as TiN or ZrN can be substitutes for conventional metals in the visible frequencies. In this paper we provide the details of fabrication and characterization of these new materials and discuss their suitability for a number of metamaterial and plasmonic applications.
\end{abstract}

\ocis{(160.3918) Metamaterials, (160.4236) Nanomaterials, (250.5403) Plasmonics,  (310.6860) Thin films, optical properties }


\section{Introduction}
The introduction of new materials into the realm of plasmonics and metamaterials (MMs) is expanding the application domain of feasible devices \cite{science_boltasseva}. Recent demonstrations of superlensing in the mid-IR \cite{perovskite_superlens}, semiconductor plasmonic quantum dots \cite{LSPR_QDs_Alivisatos} and an epsilon-near-zero (ENZ) light funnel \cite{ENZfunnel_arxiv} are but a few examples. The integration of new materials not only opens up possibilities for new devices, but it also significantly improves the performance of many existing MM and plasmonic devices. One of the most important challenges in the fields of plasmonics and MMs is the high loss in the metallic components of a device. New plasmonic materials have the promise of overcoming this major bottle-neck and enabling high-performance devices. Also, new plasmonic materials allow greater flexibility in the design of a device owing to the moderate magnitude of the real part of permittivity in such materials. On the contrary, metals such as gold and silver have very large negative real permittivities in the near-IR and visible ranges, which is a major obstacle in the design and fabrication of efficient devices. Alternative plasmonic materials have two other major advantages: they can exhibit tunable optical properties \cite{switchingTCOs_atwater}, and they can be compatible with standard fabrication and integration procedures \cite{nitride_gate}. Clearly, alternative plasmonic materials have significant advantages over conventional metals for plasmonic and metamaterial designs.
\par
Alternative plasmonic materials in the near-IR and visible ranges can be classified into categories such as semiconductor-based \cite{APM_LPR}, intermetallics \cite{noble-transMetal_alloys}, ceramics \cite{naik_cleo2011} and organic materials \cite{noginov_organicsCLEO2011}. In this article, we focus on inorganic ceramic materials: semiconductor-based oxides and transition-metal nitrides. These materials have advantages over the other types, since oxides enable low-loss all-semiconductor based plasmonic and MM devices in the near-IR, while metal-nitrides are CMOS compatible and provide alternatives to gold and silver in the visible frequencies. Here, we describe our fabrication methods for these materials and study their optical properties in the context of plasmonic and MM applications. A brief discussion on the suitability of these new materials for various plasmonic and MM devices is also presented. 

\section{Processing and characterization}
Whether we realize it or not, oxide plasmonic materials are a familiar element in everyday life since transparent conducting oxides (TCOs) are regularly used in liquid-crystal displays. These materials can exhibit metallic properties in the near-IR when heavily doped \cite{AZO_RRL}. Doping produces high carrier concentration ($N>10^{20}\, \mbox{cm}^{-3}$) which results in large plasma frequency ($\omega_p \propto \sqrt{N}$). A large plasma frequency results in Drude-metal-like optical properties \cite{drude_ashcroft}. The following equation describes the Drude response of such degenerately doped semiconductors.

\begin{equation}
\epsilon=\epsilon'+i\epsilon''=\epsilon_{\infty}-\frac{\omega_p^2}{\omega(\omega+i\gamma)}
\end{equation}

$\epsilon_{\infty}$ is due to the screening effect of bound electrons in the material and can be considered as a constant in the frequency range of interest. $\gamma$ is the Drude-relaxation rate or damping co-efficient of free carriers. This term signifies the optical losses incurred in the material. In order to have negative $\epsilon'$ in the optical range, $\omega_p$ must be large and $\epsilon_{\infty}$ must be small. Lower losses require smaller $\gamma$. Heavy doping of $10^{21}\,\mbox{cm}^{-3}$ poses problems due to solid-solubility limits in many semiconductors. However, oxide semiconductors such as zinc oxide and indium oxide overcome this problem \cite{SolidSolubility_yoon} and can be heavily doped to be metallic substitutes in the near-IR. Further increases of the carrier concentration to about $10^{22}\,\mbox{cm}^{-3}$ are required to make metal substitutes in the visible range; at those concentrations, oxides also suffer from solid-solubility limits. However, transition-metal nitrides can posses such high carrier concentrations and therefore can exhibit metallic properties at visible frequencies. In this work, we have studied the optical properties of aluminum-doped zinc oxide (AZO), gallium-doped zinc oxide (GZO), indium-tin-oxide (ITO) and nitrides of titanium, tantalum, hafnium and zirconium.

\subsection{Transparent conducting oxides}
\begin{figure}[htbp]
\centering
\mbox{\subfigure{\includegraphics[width=7cm]{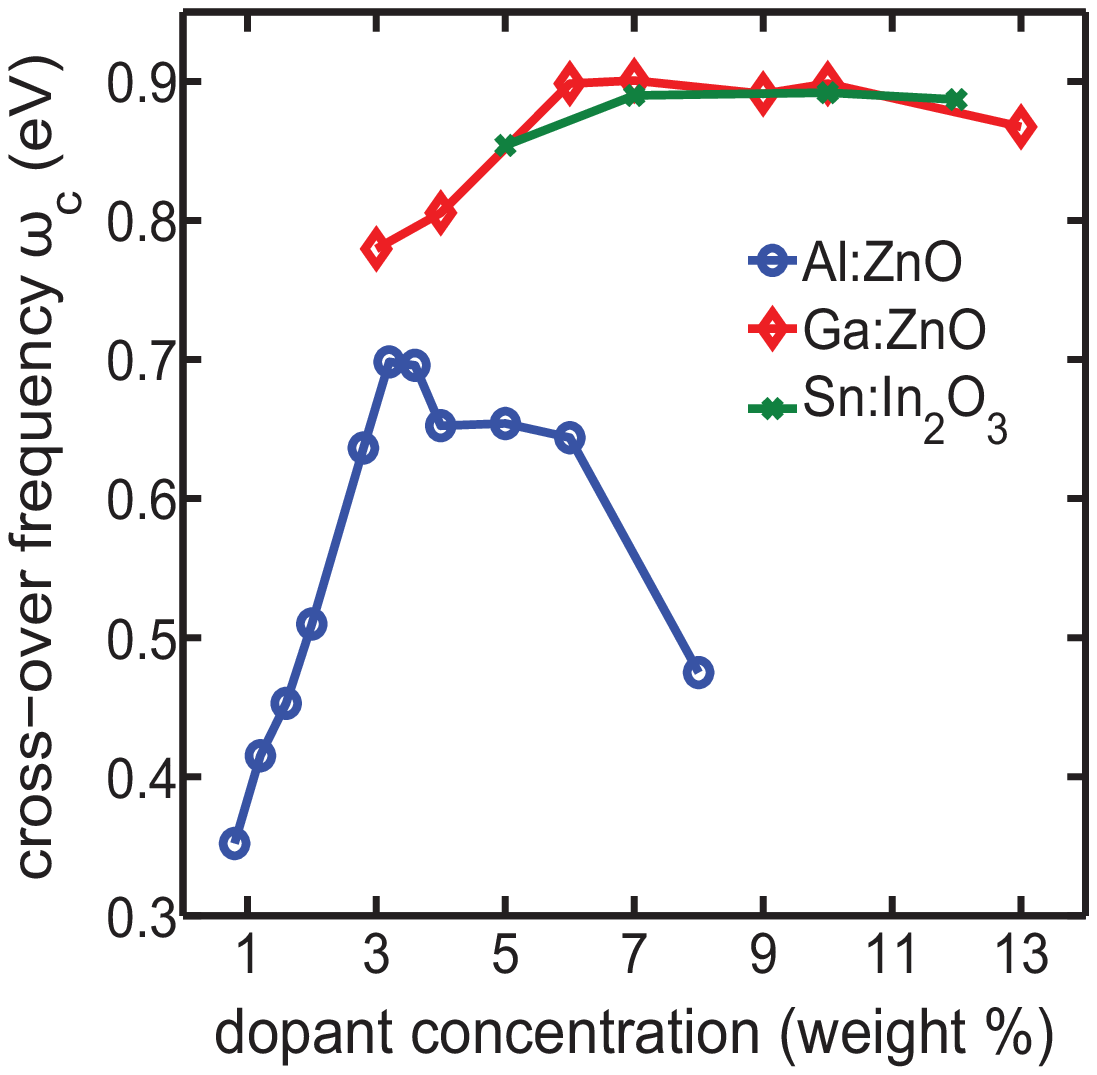}
\quad
\subfigure{\includegraphics[width=7cm]{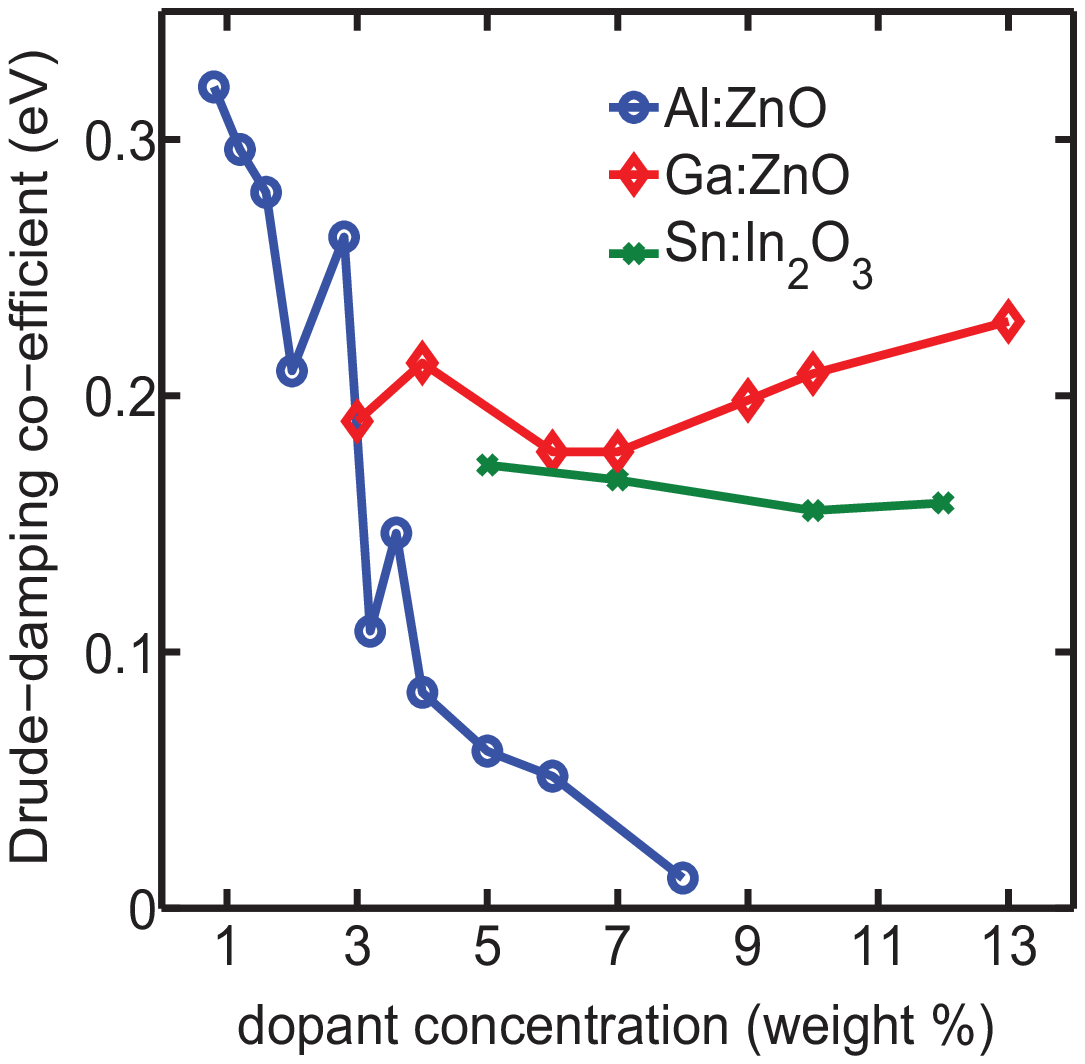}}}}
\caption{Left panel: Cross-over frequency (frequency at which real permittivity crosses zero) of Al:ZnO, ITO and Ga:ZnO films as a function of dopant concentration. Right panel: Drude-damping coefficient ($\gamma$) vs. dopant concentration. The films were deposited at 100 $^0$C (AZO and ITO) and 50 $^0$C (GZO) with oxygen partial pressures of 0.4 mTorr. The ablation energy was about 2 J/cm$^2$.}
\label{fig1}
\end{figure}

Thin films of TCOs can be deposited by many physical-vapor and chemical-vapor deposition techniques. Highly conductive TCO films can be produced by techniques such as sputtering and pulsed-laser-deposition (PLD). In our oxide-film studies, we have employed PLD (PVD Products, Inc.) with a KrF excimer laser (Lambda Physik GmbH) at a wavelength of 248 nm for source material ablation. The chosen ablation targets were $\mbox{Ga}_2\mbox{O}_3$ and ZnO for GZO, $\mbox{Al}_2\mbox{O}_3$ and ZnO for AZO, and $\mbox{In}_2\mbox{O}_3$ and $\mbox{SnO}_2$ for ITO. The targets were purchased from the Kurt J. Lesker Corp. with purities of 99.99\% or higher. The required composition of the deposited film was achieved by alternating the laser ablation over two different targets with an appropriate number of pulses on each target. A single cycle consisting of a few laser pulses on each target was repeated mnay times until the desired film thickness was achieved. The number of pulses in each cycle was designed to be small enough so that the effective layer thickness deposited in a single cycle would be less than a few atomic layers. This ensured a homogeneous mixture of the constituent materials in the final film. All the films were grown in an oxygen ambient with an oxygen partial pressure of 0.4 mTorr (0.053 Pa)  or lower. The substrate was heated to temperatures around 50-100 $^0$C during deposition. The optimization curves for GZO, ITO and AZO are shown in Fig. \ref{fig1}. The optical characterization of the thin films was performed using a spectroscopic ellipsometer (V-VASE, J. A. Woollam). The dielectric function was retrieved by fitting a Drude+Lorentz oscillator model to the ellipsometry data.

The cross-over frequency ($\omega_c$) is defined as the frequency at which the real permittivity of the material crosses zero. Since $\omega_c$ is directly proportional to plasma frequency ($\omega_p$) and $\omega_p$ is proportional to the square of carrier concentration, Fig. \ref{fig1} depicts the carrier concentration trend as a function of film composition. While the highest cross-over frequency achieved for AZO is about 0.7 eV, the same for GZO and ITO is around 0.9 eV. The substrate temperature and oxygen partial pressure during deposition of these films played significant roles in achieving the highest possible carrier concentration. The optimum values were found to be close to the parameters used in the data for Fig. \ref{fig1}.
\par
\begin{figure}[htbp]
\centering
\mbox{\subfigure{\includegraphics[width=7cm]{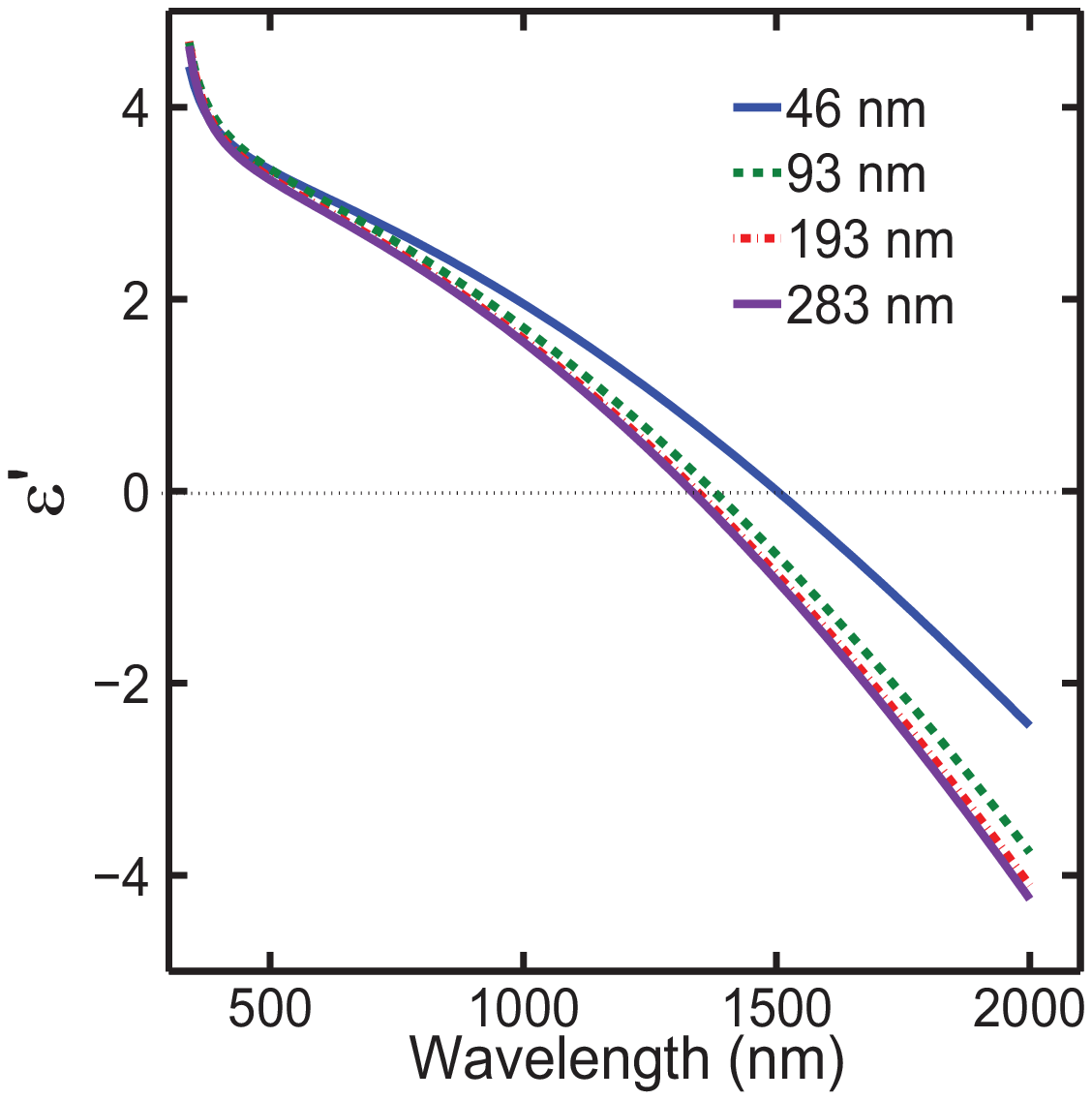}
\quad
\subfigure{\includegraphics[width=7cm]{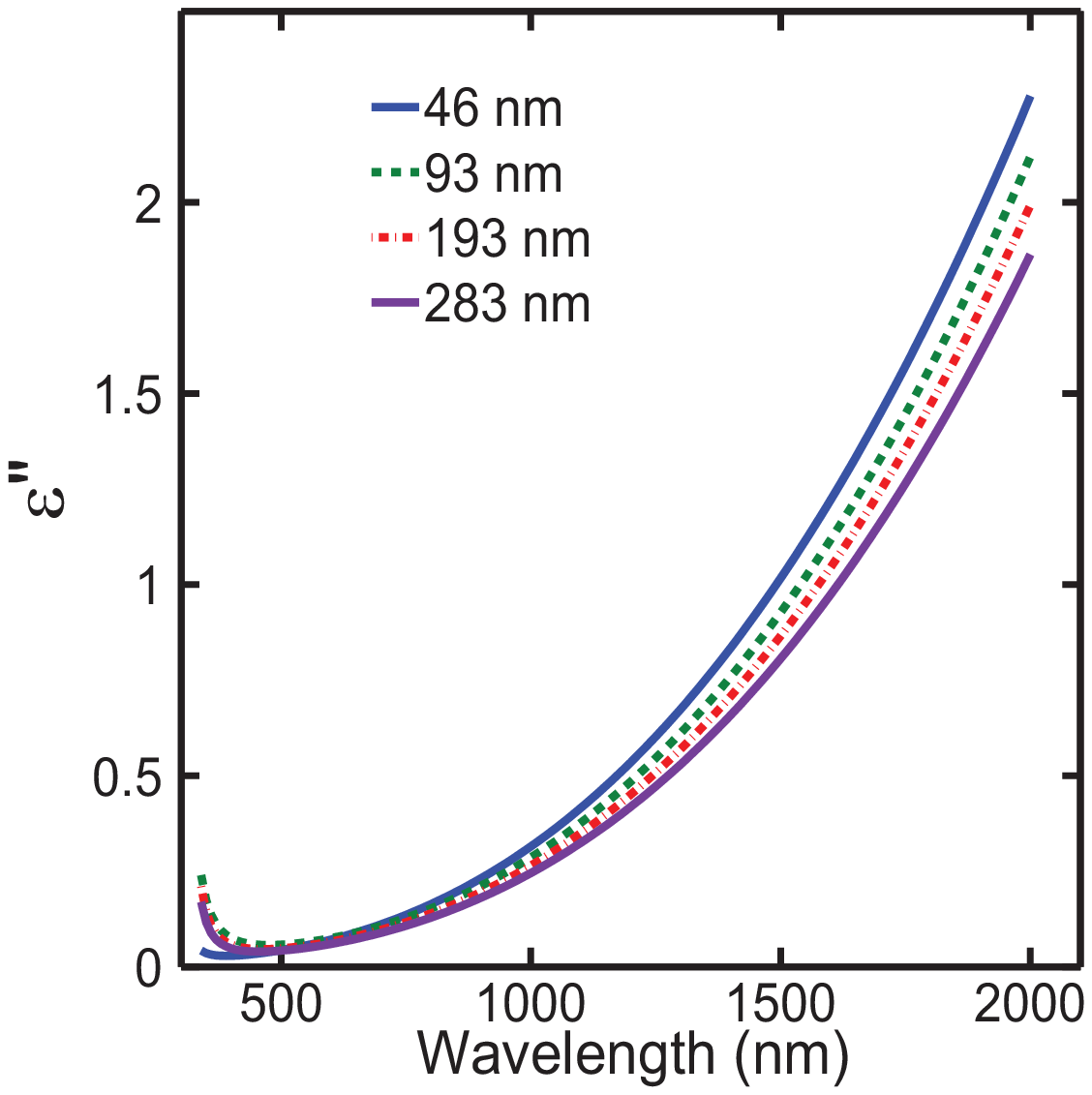}}}}
\caption{Optical properties of GZO thin films with different thicknesses deposited on glass substrates. The films were deposited under identical conditions except for the duration of deposition.}
\label{fig2}
\end{figure}

Owing to their non-stoichiometric nature, TCO films are known to exhibit thickness-dependent properties \cite{thickness_TCOs,kim2006thickness} This is because the interface with the substrate can have many carrier trap states, and hence the net carrier concentration in the film depends on the thickness (volume to surface ratio) of the film. In many nanoplasmonic devices, thin film structures are used as building blocks, and therefore it is necessary to understand how the optical properties of TCO thin films depend on their thickness. We have studied the thickness-dependent optical properties of GZO thin films on glass substrates. Figure \ref{fig2} shows the dielectric function of GZO films with different thicknesses. The films with thicknesses greater than about 50 nm exhibit very little thickness dependence in their optical properties.

\begin{figure}[htbp]
\centering
\mbox{\subfigure{\includegraphics[width=7cm]{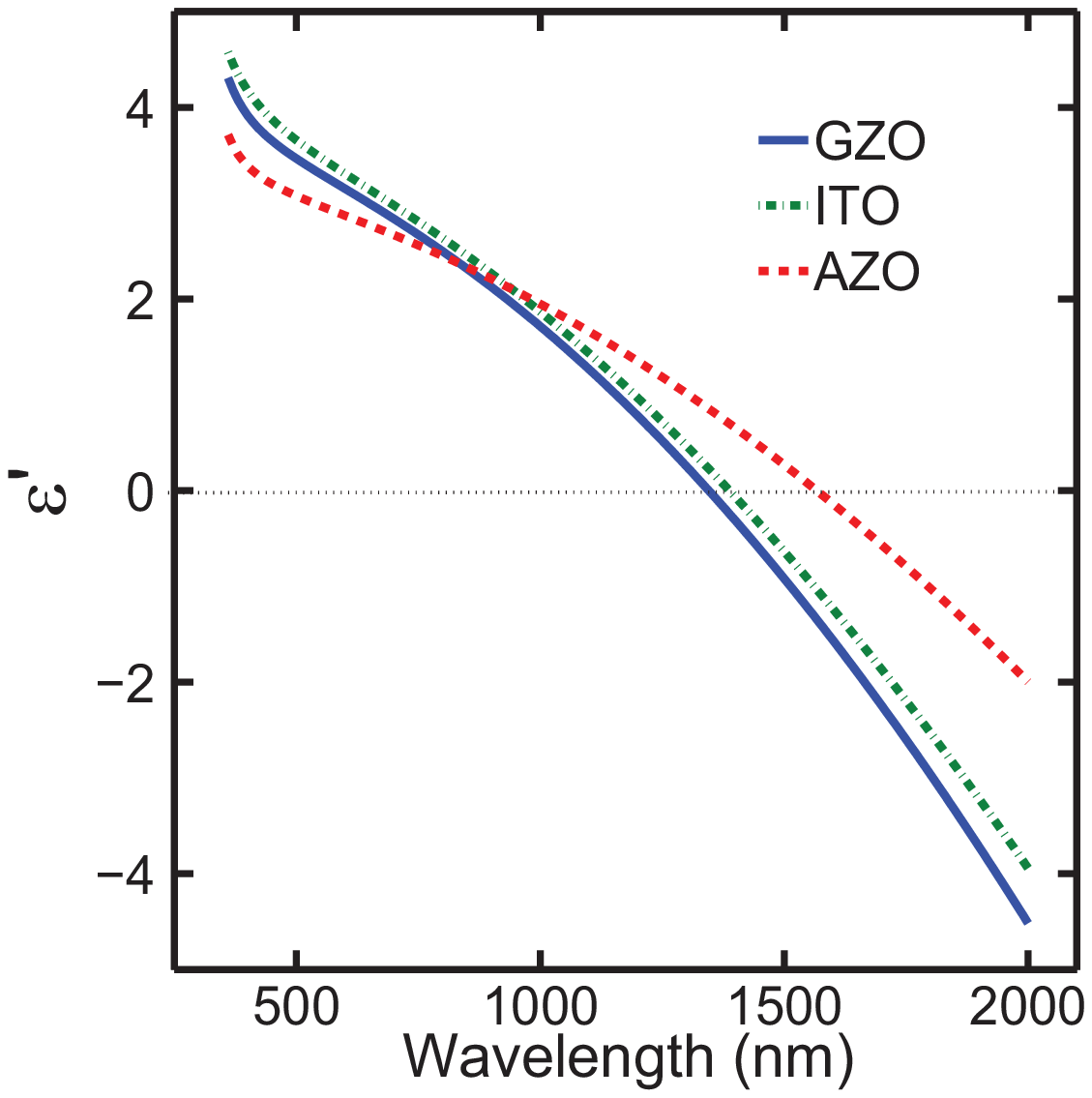}
\quad
\subfigure{\includegraphics[width=7cm]{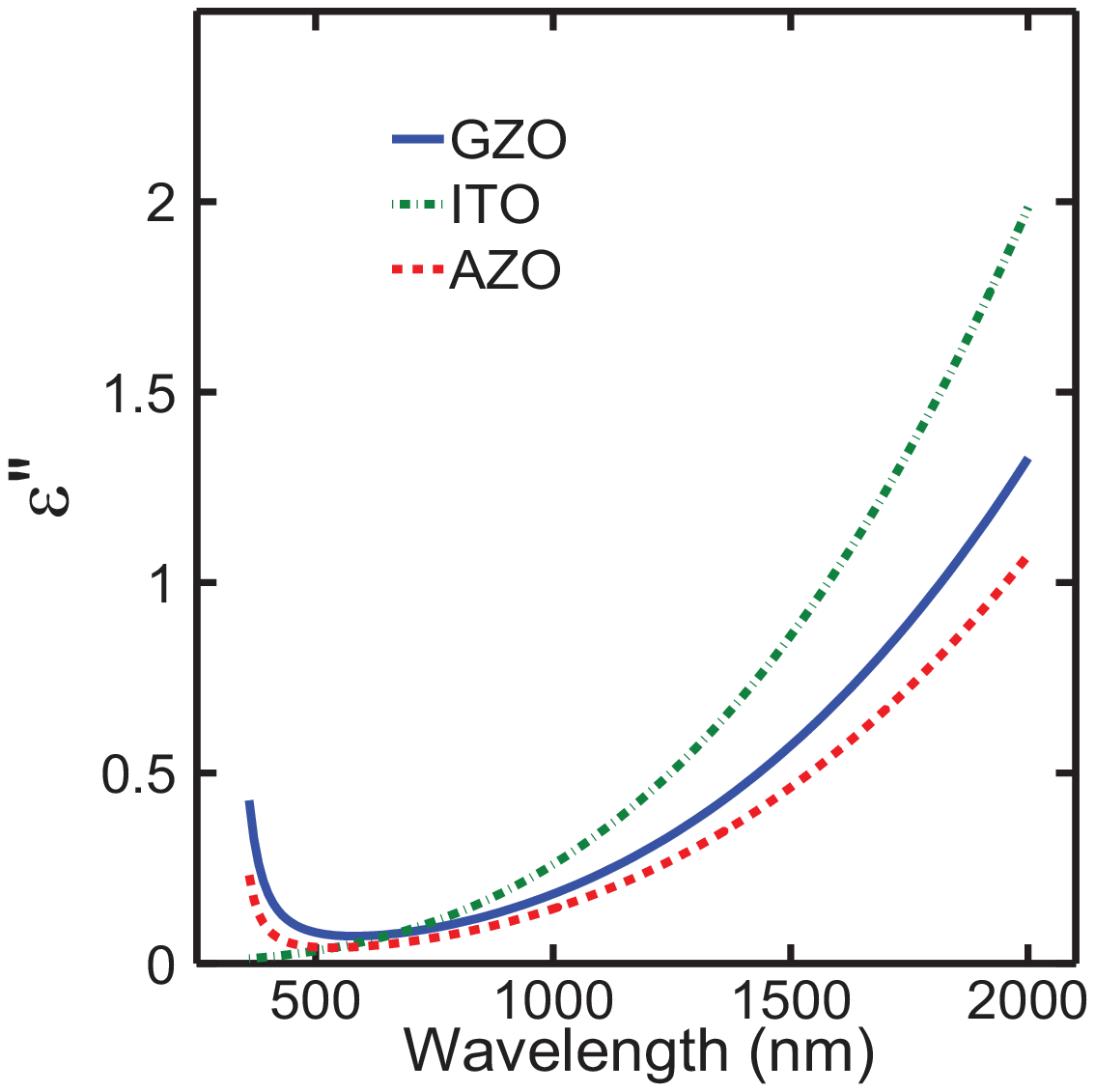}}}}
\caption{Comparison of the optical properties of pulsed laser deposited TCO films with the smallest cross-over wavelengths. The films were deposited onto glass substrates at 100 $^0C$ (AZO and ITO) and 50 $^0$C (GZO) with oxygen partial pressures of 0.4 mTorr.}
\label{fig3}
\end{figure}

As a comparison of the three different TCOs considered in our study, in Fig. \ref{fig3} we plot the optical properties of AZO, GZO and ITO films. The plots correspond to the TCO films with the lowest cross-over wavelengths and lowest losses. Notably, AZO offers the lowest Drude damping, but it also has the lowest $\omega_c$ (and hence the longest cross-over wavelength). GZO and ITO can produce cross-over wavelength as low as 1.2 $\mu$m . However, Drude damping in GZO is slightly higher than that in AZO and lower than that in ITO.

\subsection{Transition metal nitrides}
Many of the transition-metal nitrides are known to exhibit metallic properties in the visible frequencies \cite{optprop_nitrides}. Their optical properties largely depend on the deposition conditions because of their non-stoichiometric nature. As a general trend, metal-rich films are metallic and nitrogen-rich films are dielectric. As an exception, TiN is metallic both when metal-rich and when nitrogen-rich. However, the carrier concentration declines from metal-rich to nitrogen-rich films. The nitride films are also sensitive to the substrate because they show textured growth on some specific substrates. For example, c-sapphire  and MgO substrates can be low lattice-mismatched substrates \cite{epiTiN_sapphire,epiTiN_MgO}, promoting layer-by-layer crystalline growth. On the other hand, glass as a substrate produces polycrystalline or amorphous films, which exhibit quite different properties than single-crystal films.
\par
\begin{figure}[htbp]
\centering
\mbox{\subfigure{\includegraphics[width=7cm]{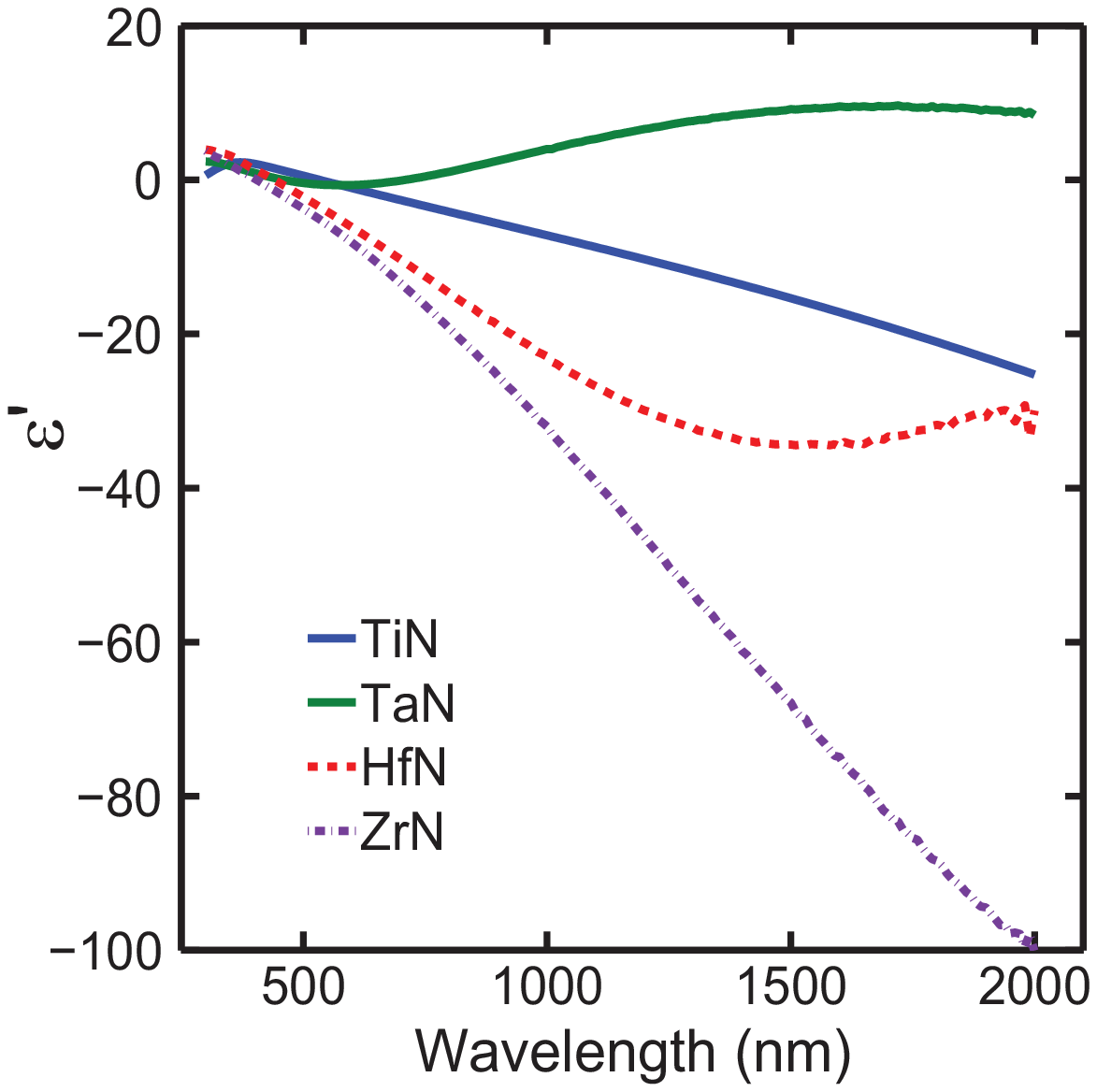}
\quad
\subfigure{\includegraphics[width=7cm]{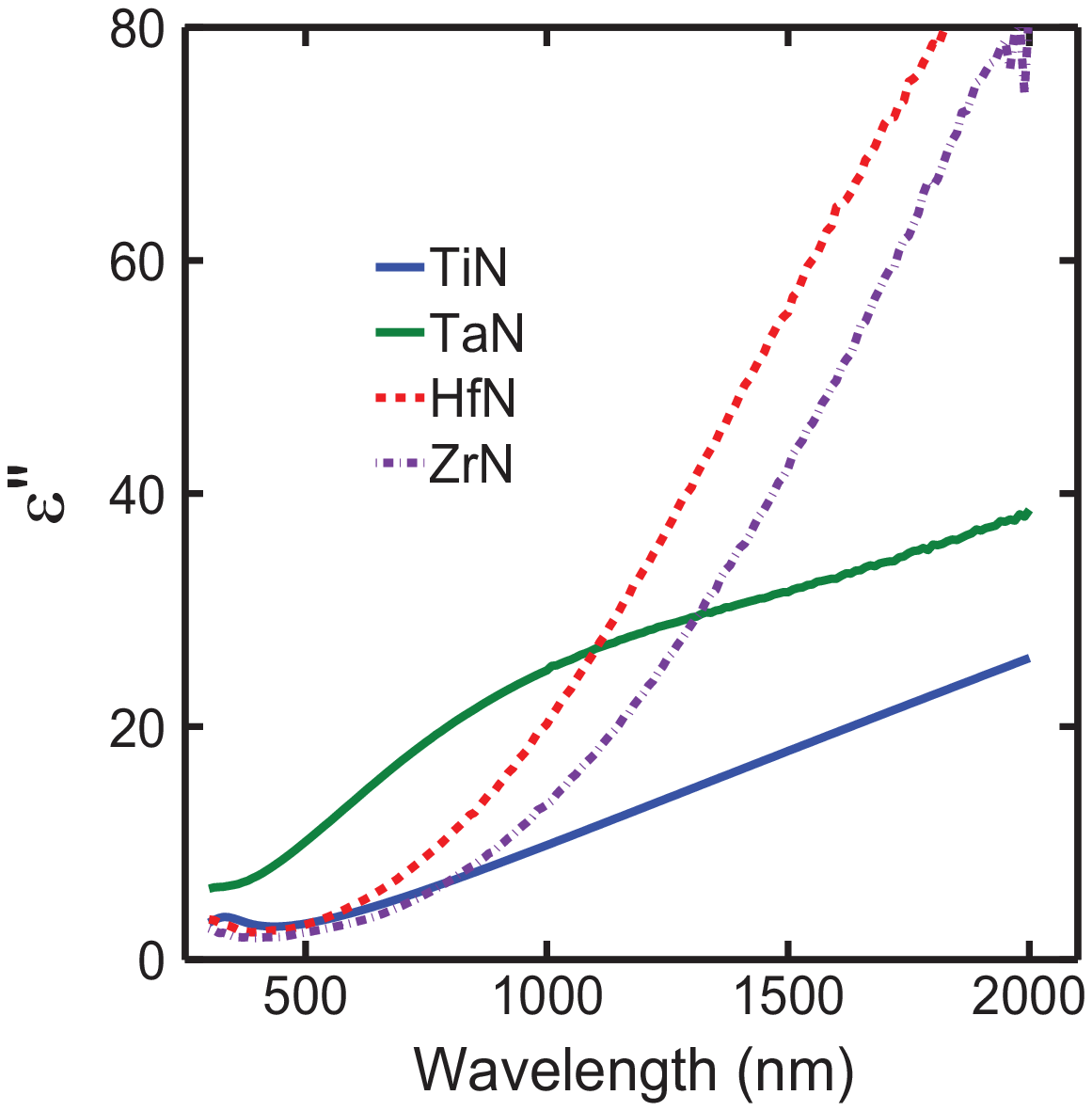}}}}
\caption{Dielectric function retrieved from spectroscopic ellipsometry measurements on thin films of TiN, TaN, HfN and ZrN deposited on c-sapphire. The substrate temperature during deposition was 800 $^0$C, and the chamber pressure was 5 mTorr. The flow ratios of $\mbox{N}_2$:Ar were 6 sccm: 4sccm for TaN and TiN films and 2 sccm: 8 sccm for the rest. The sputter power was 150 W for Ta and Hf targets and 200 W for the rest.}
\label{fig4}
\end{figure}

In our study of nitrides for plasmonic applications, we have employed DC reactive sputtering to deposit thin films of metal nitrides. Pure metal targets (99.995\%)  were sputtered in a nitrogen-argon ambient with partial pressures of nitrogen and argon adjusted by controlling their flow rates. The base pressure was about $4 \times 10^{-7}$ Torr or lower, and the sputtering pressure was held at 5 mTorr. The substrate was either glass or c-axis oriented sapphire and was heated to an elevated temperature during deposition. Films about 30-50 nm thick were deposited on the substrates, and their optical characterization was performed using a spectroscopic ellipsometer. The optical constants were retrieved using a Drude+Lorentz oscillator model accounting for intra-band and inter-band electronic contributions. The dielectric functions of thin films of TiN, TaN, ZrN and HfN deposited on sapphire substrates at 800 $^0$C are shown in Fig. \ref{fig4}. The nitrogen to argon flow ratio was maintained during deposition at 2 sccm :8 sccm for HfN, ZrN films and 6 sccm:4 sccm for TiN and TaN films. These conditions were found to produce nitride films that are metallic in the visible frequencies. Figure \ref{fig4} shows that TiN is metallic for wavelengths longer than about 500 nm. HfN and ZrN have cross-over wavelengths around 430 nm. TaN is metallic in the visible range only and becomes dielectric for longer wavelengths. This behavior is probably due to the nitrogen-rich deposition conditions. We note that the losses in these films are not small, partly because of the presence of inter-band transition losses. However, in the regions where inter-band losses are absent, the losses are mainly due to Drude damping. From Fig. \ref{fig4}, TiN and ZrN appear to be low-loss candidates for alternative plasmonic materials.

\begin{figure}[htbp]
\centering
\includegraphics[width=7cm]{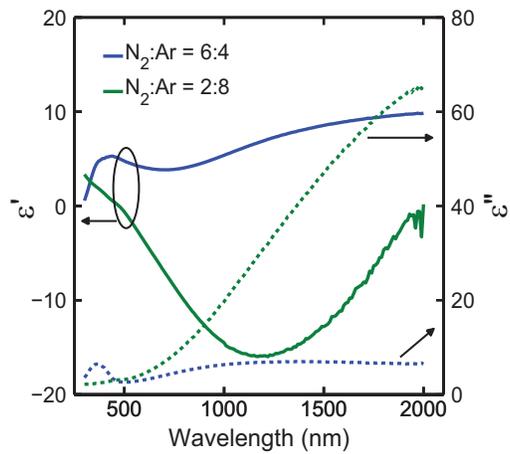}
\caption{Optical constants for ZrN films deposited on glass at different $\mbox{N}_2$:Ar flow ratios (sccm/sccm). The sputtering pressure was held at 5 mTorr and the substrate temperature during deposition was 350 $^0$C. While the ZrN film deposited in a nitrogen-rich ambient shows dielectric properties, the film deposited in a nitrogen-poor ambient shows metallic properties.}
\label{fig5}
\end{figure}

As stated previously, the optical properties of nitrides depend significantly on whether they are nitrogen-rich or metal-rich. In order to study this, ZrN thin films were deposited on glass substrates with two different nitrogen partial pressures during deposition. Figure \ref{fig5} shows the optical constants of these ZrN films. The optical properties turn from metallic for a nitrogen-poor ambient to dielectric for a nitrogen-rich ambient. This drastic change in film properties can be useful in building plasmonic devices where the properties of the material components need to be tuned or graded.

In this section the details of fabrication and their effects on the optical properties of oxides and nitrides were presented. The findings of this section shed light on the applicability of these new materials for plasmonic and metamaterial applications. The findings also help us in assessing the various advantages and disadvantages from the viewpoint of plasmonic devices as well as in terms of nanofabrication and integration. The following section discusses some of these important implications of the new class of plasmonic materials.

\section{Discussions}
\begin{figure}[tb]
\centering
\mbox{\subfigure{\includegraphics[width=7cm]{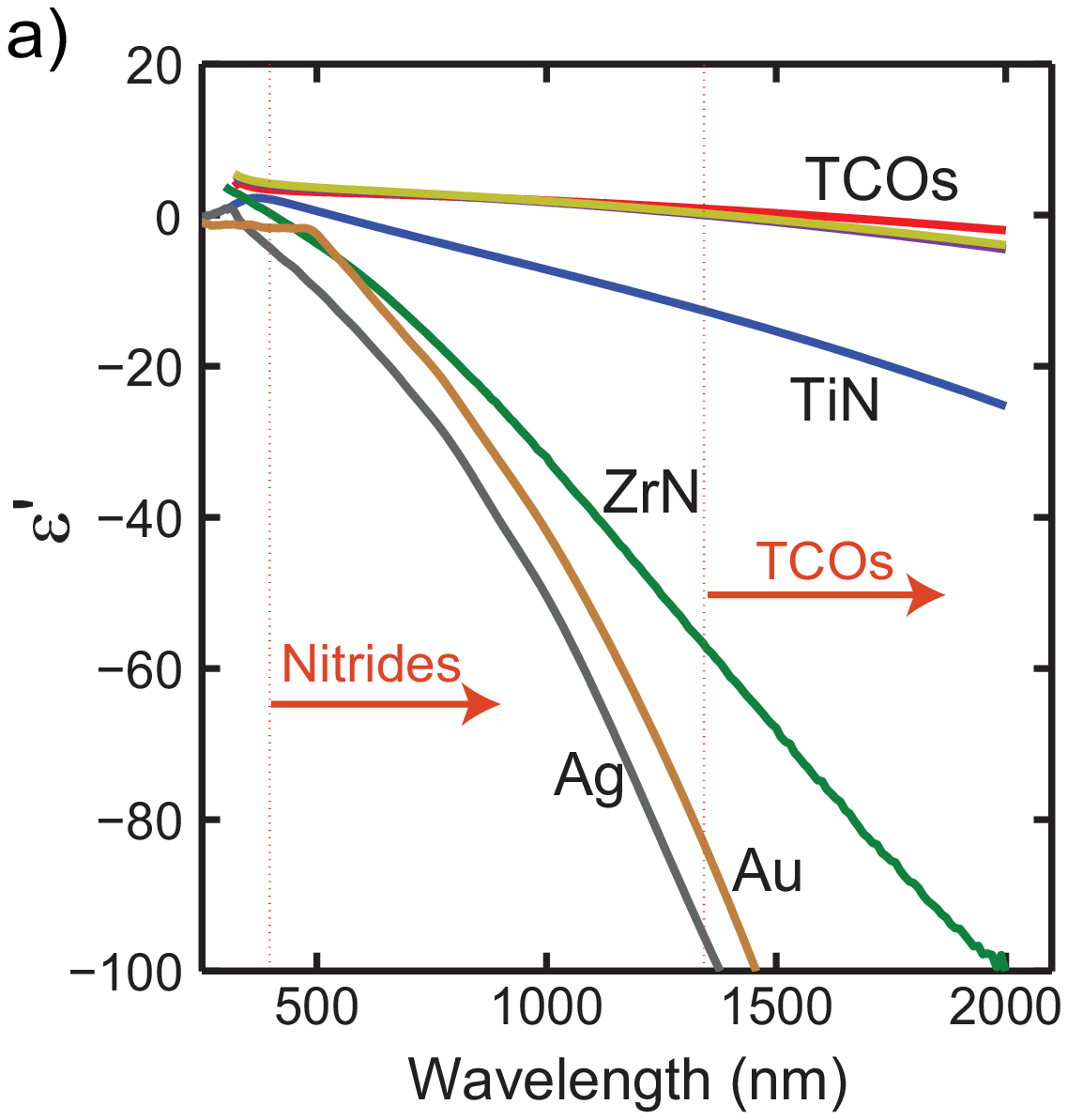}
\quad
\subfigure{\includegraphics[width=7cm]{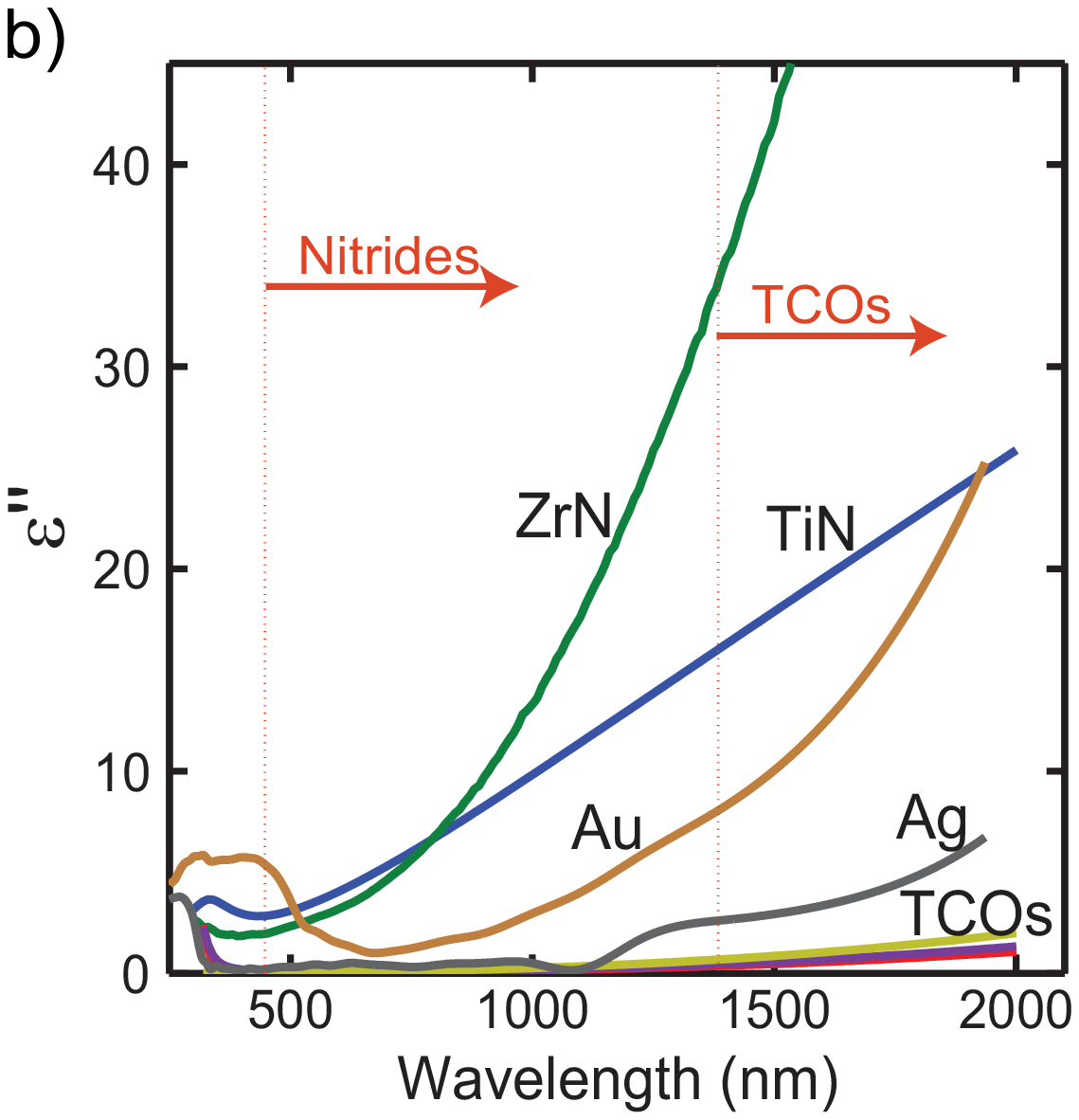}}}}
\caption{Comparison of optical properties of alternative plasmonic materials with those of conventional metals. Optical constants of low loss thin films of TiN, ZrN and TCOs (AZO, GZO and ITO) are plotted along with those of gold and silver taken from \cite{optprop_JC}. The arrows show the wavelength ranges in which nitrides and TCOs are respectively metallic. Panel a) shows that TCOs and nitrides have smaller negative permittivity values than those of metals, while b) compares losses and shows that losses in TCOs are many times smaller than those in either gold or silver. The losses in nitrides are slightly higher than metals due to inter-band transitions at the cross-over.}
\label{fig6}
\end{figure}

Novel devices made with metamaterials and transformation optics such as superlenses \cite{superlens_zhang}, hyperlenses \cite{hyperlens_zhang} and cloaks \cite{1st_cloak_smith} require special conditions to perform efficiently. One of these conditions is that the real part of permittivity of the metallic and dielectric components should be nearly the same magnitude \cite{TO_kildishev}. In the visible and near-IR frequencies, this condition is not sufficiently met by conventional metals. However, alternative plasmonic materials described in the previous section do meet this condition. In Fig. \ref{fig6}a we plot the real part of permittivity for nitrides and oxides together with those of silver and gold \cite{optprop_JC}. The arrows show the range in which nitrides and oxides are metallic. While TCOs are useful as plasmonic materials in the near-IR, nitrides are useful in the visible range. In their respective metallic ranges, both oxides and nitrides have much smaller magnitudes of real permittivity compared to those of either gold or silver.
\par 
Another important disadvantage of conventional metals is high loss. Conventional metals have very high carrier concentrations, which in turn makes their plasma frequencies very large. A large plasma frequency produces a large imaginary permittivity, which translates to large loss. Alternative plasmonic materials have lower carrier concentrations and, hence, smaller losses (see Fig. \ref{fig6}b). In particular, TCOs exhibit losses nearly five times smaller than that of the best metal (Ag) in the near-IR. This makes TCOs a good choice for many metamaterial and plasmonic applications in the near-IR and mid-IR ranges \cite{MM_TCOs,TCOs_APLNoginov}. However, the losses in transition-metal nitrides are not as low as TCOs because of larger Drude damping and inter-band losses. Comparing Drude-damping in metals and nitrides, metals have lower losses. However, upon fabrication of metallic nanostructures, the Drude-damping in metals increase many times due to imperfections such as roughness and grain-bounndary scattering \cite{Ag_drachev}. Such problems can be minimized in nitrides because they can be grown as crystalline layers on substrates like sapphire and hence, can posses very small surface roughness \cite{TiN_arxiv}. Thus, in the frequency range where inter-band losses are absent, nitrides can be good alternatives to metals.
\par
Another requirement of novel devices is tunability in the constituent material properties. Tunability can be static or dynamic depending on the application. A gradient in the properties of the constituent materials would be essential in designing many novel devices \cite{TO_kildishev,OBH_narimanov}. Conventional metals such as gold and silver cannot be tuned . Hence, they seriously limit the efficient implementation of new plasmonic and metamaterial devices. Alternative plasmonic materials do not pose such problems. Their properties are strongly dependent on the processing conditions; thus, they provide a useful method of tuning the device properties.
\par
An additional advantage of alternative plasmonic materials lies in their ease of fabrication and integration. Fabrication techniques such as chemical vapor deposition, atomic layer deposition and molecular beam epitaxy can be employed to produce oxide and nitride films of high quality. Furthermore, employing lattice-matched systems allows hetero-epitaxy, which could produce high-performance, monolithic devices. For example, AZO and GZO on sapphire or ZnO substrates could result in superlattice/monolithic devices \cite{epiAZO,epiGZO,AZO_LZOpedarnig}. Similarly, nitrides have good lattice match with sapphire or MgO substrates \cite{epiTiN_sapphire,epiTiN_MgO}. It is important to note that conventional metals such as gold and silver are not compatible with standard CMOS processes.  However, this limitation can be completely overcome by alternative plasmonic materials such as TaN and TiN, which are CMOS-compatible and have been shown to be useful as gate metals in CMOS transistors \cite{nitride_gate}.

As a summary, alternative plasmonic materials offer many advantages over conventional metals for plasmonic and metamaterial devices. Each material system considered here has its own set of advantages for a specific class of devices operating in a particular part of the optical spectrum. An appropriate choice from the subset of materials presented here could enable high-performance devices in the optical frequency range.

\section{Conclusion}
Alternative plasmonic materials have many advantages when replacing conventional metals for plasmonic and metamaterial applications. However, the properties of these alternative materials are sensitive to deposition conditions and require optimization. Optimized films of transparent conducting oxides (TCOs) such as AZO, GZO and ITO are low-loss alternatives to metals in the near-IR, where they could enable high-performance devices such as hyperlenses and superlenses. In the visible range, nitrides can be alternatives to metals although nitrides have losses slightly higher than conventional metals. Nitrides also offer many fabrication and integration advantages over conventional materials such as gold and silver.

\section*{Acknowledgments}
We thank Prof. Vladimir M. Shalaev and Prof. Timothy D. Sands for their support. We thank Prof. Thomas A. Klar, Prof. Johannes Pedarnig, Prof. Peter Schaaf and Dr. Marius-Aurel Bodea for helpful discussions. This work was supported in part from ARO grant W911NF-09-1-0516.

\end{document}